\documentclass[10pt]{article}
\usepackage{amsmath}
\usepackage{epsf} 
\usepackage{graphicx}
\usepackage{amsfonts}
\usepackage{amssymb}
\begin{document}
\newcommand {\sheptitle}
{A new method of parameterisation of neutrino mass matrix through breaking of $\mu-\tau$ symmetry: Normal hierarchy }
\newcommand{\shepauthor}
{Subhankar Roy and N. Nimai Singh}
\newcommand{\shepaddress}
{Department of Physics, Gauhati University, Guwahati-781014, India\\E-mail:meetsubhankar@gmail.com }
\newcommand{\shepabstract}
{The present observational data on neutrino oscillation strongly supports the deviation from Tri-bimaximal mixing (TBM) and nonzero value of $\theta_{13}$. In the first part of the present work, the $\mu-\tau$ symmetry of the neutrino mass matrix is perturbed at its minimal level in order to produce deviation from TBM, which includes nonzero value of $\theta_{13}$ and maximal condition of $\tan^{2}\theta_{23}=1$. The parameterisation of neutrino mass matrix which describes Normal hierarchy (N.H), has been addressed with minimum number of independent parameters, out of which two parameters $\eta$ and $\alpha$ control $\theta_{12}$ and $\theta_{13}$ respectively, without any interference with mass eigenvalues. This model is found to be flexible enough to adjust itself with the changing precise experimental results. In the second part, the deviation from maximal condition $\tan^{2}\theta_{23}=1$, along with the nonzero value of $\theta_{13 }$, has been implemented with the introduction of a perturbating matrix $M_{s}$ which breaks the $\mu-\tau$ symmetric mass matrix.\\ 
\\
PACS numbers: 14.60.Pq, 12.15.Ff, 13.40.Em\\
\\
Keywords: Normal Hierarchy, Tri-bimaximal mixing, neutrino masses, $\mu-\tau$ symmetry.

} 

\begin{titlepage}
\begin{flushright}
\end{flushright}
\begin{center}
{\large{\bf\sheptitle}}
\bigskip \\
\shepauthor
\\
\mbox{}
\\
{\it \shepaddress}
\\
\vspace{.5in}
{\bf Abstract} \bigskip \end{center}\setcounter{page}{0}
\shepabstract
\end{titlepage}

\newpage
\section{Introduction}
The recent discovery that the reactor angle $ \theta_{13}$ is not only nonzero but relatively large, by the Daya Bay and RENO experiments [1, 2], has a  significant impact on the existing neutrino mass models. The global bestfit values of the neutrino oscillation parameters like $\tan^{2}\theta_{12}$ and $\sin^{2}\theta_{13}$ are 0.047 and 0.026 (N.H) respectively [3]. The data on mass-squraed differences are very precise  and the Dirac delta phase $\delta_{cp}$ is still in the dark. We have taken $\delta_{cp}=0$ throughout the calculations and assumed that there is no sterile neutrino.
 
 Many theories predict that the atmospheric mixing angle, $\tan^{2}\theta_{23}$ must depart from maximality condition [4, 5, 6] when $\mu-\tau$ symmetry is broken in order to produce nonzero $\theta_{13}$. Two possibilities are there in conection with deviation from TBM along with the generation of nonzero $\theta_{13}$ and they are either with $\theta_{23}=\pi/4$ or with $\theta_{23}\neq\pi/4$. From theoretical point of view, the problem can be addressed either by perturbing the $\mu-\tau$ symmetry of the mass matrix to generate the desired results [7] or by starting from a new PMNS mixing matrix which can produce the present experimental results [8].  

The pattern of the absolute neutrino masses whether Normal hierarchy (NH) or Inverted hierarchy (IH) [9], is still an open question. Besides, the status of quasidegenerate (QD) model of neutrino mass pattern is not yet completely ruled out. The $\mu-\tau$ symmetry is capable of producing TBM mixing and deviation as well [10,11]. In this present work a new method of parameterisation of neutrino mass matrix is presented with the hope to perturb the $\mu-\tau$ symmetry to get the desired result. In this method, the parameterisation of $\mu-\tau$ symmetric matrix can be done with three independent parameters i.e., $M_{\mu\tau}$ $(\xi,\eta,m_{0})$ where solar mixing angle $\theta_{12}$ depends upon only one parameter $\eta$ where the mass eigenvalues $m_{i}$ depend upon the rest i.e., $(\xi,m_{0})$. Such a procedure has the advantage to delink the predictions on mass eigenvalues and mixing angles.

The matrix $M_{\mu\tau}(\xi,\eta,m_{0})$ is again perturbed to break the $\mu-\tau$ symmetry by adding an extra parameter $\alpha$ which controls the prediction of reactor angle $\theta_{13}=\theta_{13}(\alpha)$, keeping $\tan^{2}\theta_{23}=1$ as maximal condition. The possibility to obtain nonzero value of $\theta_{13}$ along-with $\tan^{2}\theta_{23}\neq1$, is examined with the help of a perturbing mass matrix $M_{s}$, i.e $M_{\mu\tau}+M_{s}\longrightarrow M$, where $M_{\mu\tau}$ is fixed at TBM.

The paper is organized in the following manner. In Section 2, we discuss the general conditions for parameterisation  related with exact $\mu-\tau$ symmetric and general symmetric mass matrices.
 In Section 3, we also present the procedure regarding parameterisation of the mass matrix under N.H, both for the cases with $\theta_{13}=0$ and $\theta_{13}\neq0$ (with $\theta_{23}=\pi/4$) and the numerical results.
 In Section 4, we introduce a new perturbative matrix $M_{s}$ in order to perturb the $M_{\mu\tau}$ satisfying TBM mixing. The variation of all the observed parameters are studied under a single parameter.
We present a brief summary and discussion in Section 5. 

\section{Exact $\mu-\tau$ symmetry and general symmetric mass matrices}
The $\mu-\tau$ symmetry is a very beautiful symmetry which provides a good control over the choice of the solar mixing angle $\theta_{12}$. Tribimaximal mixing (TBM) which is associated with certain symmetry groups like $A_{4}$ and $ S_{4}$ etc. is an outcome of the exact $\mu-\tau$ symmetry. But in order to introduce nonzero value of $\theta_{13}$, this symmetry has to be broken down. Under the assumption that neutrinos are Majorana type, we get a freedom to preserve the symmetric nature of the mass matrix. A $\mu-\tau$ symmetric mass matrix 
\begin{eqnarray}
M_{\mu\tau} & = & \begin{pmatrix}
A & B  & B \\ 
B & C & D \\ 
B & D & C
\end{pmatrix}
\end{eqnarray}
leads to
\begin{eqnarray}
\tan2\theta_{12} & = & \frac{2\sqrt{2} B}{A-C-D}  .
\end{eqnarray}  
Under the condition that $\theta_{13} = 0$, $\theta_{23}=\pi/4$ the corresponding PMNS mixing matrix $U$ becomes,
\begin{eqnarray}
U=\begin{pmatrix}
\cos\theta_{12}  & -\sin\theta_{12} & 0  \\
  \frac{1}{\sqrt{2}} \sin\theta_{12}& \frac{1}{\sqrt{2}} \cos\theta_{12} & -\frac{1}{\sqrt{2}}   \\
  \frac{1}{\sqrt{2}} \sin\theta_{12} & \frac{1}{\sqrt{2}} \cos\theta_{12}  &  \frac{1}{\sqrt{2}} \\
  \end{pmatrix}.
\end{eqnarray}\\
When $M_{\mu\tau}$ in Eq (1) is broken down to a general matrix which can impart a nonzero $\theta_{13}$,
\begin{eqnarray}
M = \begin{pmatrix}
A' & B'  & B'' \\ 
B' & C' & D' \\ 
B'' & D', & C''
\end{pmatrix},
\end{eqnarray}
then the PMNS mixing matrix $U$, takes the following form,
\begin{eqnarray}
U = \begin{pmatrix}
 \cos\theta_{12} \cos\theta_{13}  &-\sin\theta_{12} \cos\theta_{13}  &\sin\theta_{13}   \\
 \frac{1}{\sqrt{2}} (\sin\theta_{12} + \cos\theta_{12} \sin\theta_{13})  & \frac{1}{\sqrt{2}} (\cos\theta_{12} - \sin\theta_{12} \sin\theta_{13})  &   \frac{-1}{\sqrt{2}} \cos\theta_{13} \\
 \frac{1}{\sqrt{2}} (\sin\theta_{12} - \cos\theta_{12} \sin\theta_{13})   & \frac{1}{\sqrt{2}} (\cos\theta_{12} + \sin\theta_{12} \sin\theta_{13}) &   \frac{1}{\sqrt{2}} \cos\theta_{13}  \\
\end{pmatrix}.
\end{eqnarray}
On diagonalizing $M$ in Eq.(4), with this new $U$, $U^{T} M U \longrightarrow M_{d}$, we arrive at the following two important conditions, under the fullfillment of which the complete diagonalization is possible. They are,
\begin{footnotesize}
\begin{eqnarray}
\tan2\theta_{12} = \frac{-4 (\sqrt{2}(B'+B'')\cos\theta_{13}+(C'-C'')\sin\theta_{13})}{-2 A'+C'+C''+6 D'+(-2 A'+C'+C''-2 D')\cos 2\theta_{13} + 2 \sqrt{2} (B'+B'') \sin 2 \theta_{13} }
\end{eqnarray}\\
\end{footnotesize}
and
\begin{eqnarray}
\cot\theta_{12}= \frac{2 ( (-C'+C'')\cos \theta_{13} + (B' + B'') \sin \theta_{13} )}{2 \sqrt{2}(B'-B'') \cos\theta _{13} + (-2 A'+ C'+C''-2 D') \sin 2\theta_{13}}.
\end{eqnarray}
These two equations involving $\theta_{12}$ and $\theta_{13}$ add little complicacy in the process of parametrisation. This can be emphasised that atmospheric mixing angle is taken as maximal i.e, $\theta_{23}=\pi/4$.
\section{Building the  Neutrino mass model under Normal Hierarchy }
Normal hierarchy is the case when we take the absolute masses of the three neutrinos in the order  $m_{1}$\hspace{1mm}$<$\hspace{1mm}$ m_{2}$\hspace{1mm}$<$ $m_{3}$. The mass of $m_{1}$ is considered to be very small in comparison to $m_{2}$ and $m_{3}$ and can be taken to be nearly zero.
\subsection{Formalism}
For parameterisation of the mass matrices it is always kept in mind that when the perturbating effect is to be nullified, we must arrive at the original $\mu- \tau$ symmetric structure. $M_{\mu\tau}$ attracts special attention in our work. The parametrisation of the mass matrices, both for the cases when $\theta_{13} = 0 $ and $\theta_{13}\neq 0$, are adderessed with equal footing.
The eigenvalues of the mass matrix, give the three absolute masses $m_{1}, m_{2}$ and $ m_{3}$ and from the mass matrix itself we can generate $U$, the PMNS mixing matrix which in turn provides us the information of solar and reactor mixing angles, $\theta_{12}$ and $\theta_{13}$ respectively. Hence in order to parameterize the neutrino mass matrix we shall be requiring four (if $\theta_{13} = 0$) or five (if $\theta_{13} \neq 0$) independent parameters. We have given adequate attention in excluding any additional parameter or constant from our model which allows only five parameters at the most. 
In the process of parameterisation we always try to satisfy strongly the following two objectives:
\begin{itemize}
\item The eigenvalues of the mass matrix must have a simple form and should not involve any parameter that controls $\theta_{12}$ or $\theta_{13}$.
\item The expression of $\tan 2 \theta_{12}$ must involve a single guiding parameter at least for the case when $\theta_{13}=0$.
\end{itemize}
It is to be mentioned that our process of parameterisation assumes Dirac phase $\delta_{CP}$ to be zero all the time and is bound to produce $\tan^{2}\theta_{23}=1$. 
\subsubsection{Parametrisation of the mass matrix for Normal hierarchy with exact $\mu -\tau $ symmetry (when $\theta_{13}=0$)}
The matrix $M_{\mu\tau}$ in (1) has three eigenvalues as follows:
\begin{eqnarray}
& C-D ,& \\
&\frac{1}{2} (A + C + D - \sqrt{A^{2} + 8 B^{2}-2 A C + C^{2} - 2 A D + 2 C D + D^2}),&\\
&\frac{1}{2} (A + C + D + \sqrt{A^{2} + 8 B^{2}-2 A C + C^{2} - 2 A D + 2 C D + D^2}).&
\end{eqnarray}
One of the eigenvalues and hence one parameter can be normalised to unity in view of simplicity and at the end we can recover this once again. For Normal hierarchy we can take $m_{1}= 0$. Hence only two parameters $ \xi$ and $\eta$ can fulfill our purpose. On equating the first and the second eigenvalues to 1 and 0, we get 
\begin{eqnarray}
C = 1 + D \quad \text{and}\quad B = -\sqrt{\frac{A (1 + 2 D)}{2}}.
\end{eqnarray}
In order to satisfy Eq.(11), we take $D$ to be $-1/2$, $ C= 1/2$ and $ B = 0$. Thus,
\begin{eqnarray}
M_{\mu\tau} \longrightarrow M_{0} &=& \begin{pmatrix}
A & 0 & 0 \\ 
0 & \dfrac{1}{2} & -\dfrac{1}{2} \\ 
0&  -\dfrac{1}{2} & \dfrac{1}{2}
\end{pmatrix},\\
\tan 2 \theta_{12} & = & 0\nonumber
\end{eqnarray}
We consider another $\mu-\tau$ symmetric mass matrix $M_{1} $ which is added to $M_{0}$ to get full $M_{\mu\tau}$. Thus $M_{0}+ M_{1}\longrightarrow M_{\mu\tau}$. We choose,
\begin{eqnarray}
M_{1} = \begin{pmatrix}
 0& B  & B  \\ 
 B& x  & x \\ 
B & x & x
\end{pmatrix}\quad
\text{to get},\quad
M_{\mu\tau}=\begin{pmatrix}
A & B & B \\ 
B & \frac{1}{2}+x & -\frac{1}{2}+x \\ 
B & -\frac{1}{2}+x &  \frac{1}{2}+x
\end{pmatrix} .
\end{eqnarray}
Eqs.(11) and (13) give,
\begin{equation} 
B = -\sqrt{Ax},
\end{equation} 
and we get,
\begin{eqnarray} M_{\mu\tau} & = &  \begin{pmatrix}
A &-\sqrt{Ax}  &-\sqrt{Ax}  \\ 
-\sqrt{Ax} &\frac{1}{2}+x  & -\frac{1}{2}+x \\ 
-\sqrt{Ax} &-\frac{1}{2}+x  & \frac{1}{2}+x
\end{pmatrix}, \\
\tan 2 \theta_{12} & = & \frac{-2\sqrt{ 2 A x}}{A - 2 x}.
\end{eqnarray}
Let us consider, $A$ and $x$ are functions of two variables $\xi$ and $\eta$. We choose 
\begin{eqnarray}\nonumber
A = A(\xi,\eta)=A(\xi)A(\eta)\quad\text{and}\quad x =  x(\xi,\eta)=x(\xi)x(\eta).
\end{eqnarray} 
In Eq.(16), $\tan 2 \theta_{12}$ is now function of both $\xi$ and $\eta$. To make it free from $\xi$ 's, we make the following simple choice,
\begin{eqnarray}
A(\xi) = x(\xi) = \xi ,
\end{eqnarray}\\
leading to,
\begin{eqnarray}
\tan 2 \theta_{12} = \frac{-2 \sqrt{A(\eta) x(\eta)}}{A(\eta)-2 x(\eta)}=\tan 2\theta_{12}(\eta).
\end{eqnarray}
In Eq.(15), $M_{\mu\tau}$ has eigenvalues $ 1 $, $0$ and $ A(\xi,\eta) + 2 x (\xi,\eta)$,
and 
$A(\xi,\eta) + 2 x (\xi,\eta) = \xi (A(\eta)+2x(\eta))$.
Following our apriori condition, we can choose $A(\eta)+2x(\eta)=1$, so that $A(\xi,\eta) + 2 x (\xi,\eta)$ is dependent on $\xi$  only.
As square roots are involved in Eq.(15), we can have a simplest choice for $A(\eta)$, i.e $A(\eta)= \eta^{2}$.
which gives, $x(\eta)=\dfrac{1}{2} (1- \eta^{2})$.
Finally, we get\begin{eqnarray}
A = \xi \eta^{2} \quad \text{and} \quad
x = \frac{\xi}{2}(1-\eta^{2}).
\end{eqnarray}
Eqs.(15) and (16) become \begin{eqnarray}
M_{\mu\tau}=\begin{pmatrix}
 \xi \eta^{2}& - \xi \eta \sqrt{\frac{1-\eta^{2}}{2}}  & - \xi \eta \sqrt{\frac{1-\eta^{2}}{2}} \\ 
- \xi \eta \sqrt{\frac{1-\eta^{2}}{2}} & \frac{1}{2}(1 + \xi (1- \eta^{2})) & \frac{1}{2}(-1 + \xi (1- \eta^{2})) \\ 
- \xi \eta \sqrt{\frac{1-\eta^{2}}{2}} &\frac{1}{2}(-1 + \xi (1- \eta^{2}))  & \frac{1}{2}(1 + \xi (1- \eta^{2}))
\end{pmatrix} m_{0},
\end{eqnarray}
\begin{eqnarray}
\tan 2 \theta_{12} = \frac{2 \eta \sqrt{1- \eta^{2}}}{1-2 \eta^{2}}.
\end{eqnarray}
As stated earlier we have taken one of the eigenvalues to be unity, the parameter $m_{0}$ appears as a compensation for that. The eigenvalues of Eq.(20) are 
\begin{eqnarray}
0, \quad m_{0} \xi \quad \text{and} \quad m_{0}.
\end{eqnarray}
Here, the largest eigenvalue $m_{3}$ is assigned to $m_{0}$ in general.

The deviation from TBM mixing can be obtained from charged lepton corrections in the usual way [8]. The breaking of $\mu-\tau$ symmetry as a source of deviation from TBM mixing without charged lepton correction, is addressed in the subsequent subsections.
\subsubsection{Parameterisation of the mass matrix under Normal Hierarchy for $\theta_{13}\neq 0$ with broken $\mu-\tau$ symmetry.}
Here we consider that $\mu-\tau$ symmetry is broken down and the mass matrix is now a general symmetric matrix which has five unknown elements. But as we can normalize one of the eigenvalues to unity and other to zero (for NH), only three independent three independent parameters $\xi, \eta$ and $\alpha $ are required for parameterisation. The procedure of parametrisation is again outlined as follows. Consider $B=0$, then,
\begin{eqnarray}
M_{\mu\tau}\longrightarrow M_{0} = \begin{pmatrix}
 A& 0 & 0 \\ 
 0& C & D \\ 
 0& D & C
\end{pmatrix} .
\end{eqnarray}
$M_{0}$ is perturbed with $M_{1}$ which is also $\mu-\tau$ symmetric, to get a new $\mu-\tau$ symmetric matrix $ (M_{\mu\tau})_{new}$,  $M_{0} + M_{1} \longrightarrow (M_{\mu\tau})_{new}$. We take 
 \begin{eqnarray}
M_{1} = \begin{pmatrix}
a & b B  & b B \\ 
 b B &-\dfrac{a}{2}  & \dfrac{a}{2} \\ 
 b B & \dfrac{a}{2}	 & -\dfrac{a}{2}
\end{pmatrix},
(M_{\mu\tau})_{new} = \begin{pmatrix}
A + a & b B   & b B   \\ 
b B	 & C-\frac{a}{2} 	& D+\frac{a}{2} \\ 
b B	 & D+\frac{a}{2} 	& C-\frac{a}{2}
\end{pmatrix} .
\end{eqnarray}
We consider a symmetric matrix $M_{s}$ in order to perturb $(M_{\mu\tau})_{new}$. Thus
 $(M_{\mu\tau})_{new} + M_{s}\longrightarrow M :$
\begin{eqnarray}
M_s =\begin{pmatrix}
0 & -y & +y \\ 
-y & -x & 0 \\ 
+y & 0 & +x
\end{pmatrix},
\end{eqnarray}
We have chosen the elements at $2-3$ and $3-2$ positions as zero in the structure of perturbing matrix $M_{s}$ [13, 14]. This choice helps in keeping the maximality condition of $\tan^{2}\theta_{23}$.
and we get,
\begin{eqnarray}
M =\begin{pmatrix}
 A +a & b B -y & b B+y  \\ 
 b B -y& C-\frac{a}{2}-x & D+\frac{a}{2} \\ 
b B+y & D+\frac{a}{2} &  C-\frac{a}{2} + x
\end{pmatrix}.
\end{eqnarray}
All the terms $A, B, C, D, a, b, x$ and $y$ are functions of $\xi, \eta$ and $\alpha$. We have to choose $a, b, x$ and $y$ in such a way that under a choice of $\alpha = 0$, $M$ must converge to the original $\mu-\tau$ symmetric form of Eq.(20). This allows us to choose $A, B, C$ and $D$ having the same structure as in Eq.(20). Thus, we get
$ A  = \xi \eta^{2}$,
$ B  = - \xi \eta \sqrt{\frac{1-\eta^{2}}{2}}$,
 $C =  \frac{1}{2} (1 + \xi (1 - \eta ^{2}))$ and
$ D = \frac{1}{2} (-1 + \xi (1 - \eta ^{2}))$.
We have assumed the structure of $a$ and $b$ as,
$a = \alpha^{2} (1 - \xi \eta^{2})$ {and} $ b = \sqrt{1- \alpha^{2}}$.
These assumptions lead to the eigenvalues of $ M $ to attain simplest structures. Eqs.(6) and (7) are rather complicated and less useful for our purpose. The structure of $x$ and $y$ has to be worked out from the constraints on the matrix $M$. 
If $\lambda_{i = 0,1,2}$ are the eigenvalues, then in Eq.(26), $M$ must satisfy the eigenvalue equation,\begin{eqnarray}
det|M - \lambda_{i}| = 0.
\end{eqnarray}
From apriori condition, we can take $\lambda_{0} = 0$ and $\lambda_{1} = 1$.
Accordingly, from Eq.(27), we can construct two equations. They are as follows,
\begin{eqnarray}\nonumber
& & -\alpha^{4} \xi(-1 + \eta^{2})(-1+ \xi\eta^{2})+\alpha^{2}(-x^{2}+\xi(1+\eta^{2}(-1 +x^{2})))\\
& &-\xi(2\sqrt{2(1-\alpha^{2})}\eta\sqrt{1-\eta^{2}}xy+2y^{2}+\eta^{2}(x^{2}-2y^{2}))=0 ,
\end{eqnarray}
\begin{eqnarray}\nonumber
& &\alpha^{4}(1+\xi^{2}\eta^{2}-\xi(1+\eta^{2}))+(1-\xi\eta^{2})x^{2}+\alpha^{2}(-1-\xi^{2}\eta^{2}- x^{2}+
\xi(1+\eta^{2}.\\
& &\begin{tiny}
\end{tiny}(1+x^{2})))-2\sqrt{2(1-\alpha^{2})}\xi\eta\sqrt{1-\eta^{2}}xy+2(1+\xi(-1+\eta^{2}))y^{2}=0
\end{eqnarray}
Solving Eqs.(28) and (29) for $x$ and $y$ we get, $x=\alpha\xi\eta \sqrt{1-\eta^{2}}$ and $y=\alpha\sqrt{\dfrac{1-\alpha^{2}}{2}}(1-\xi\eta^{2}).$
Rearranging all the elements in (26), we get, $M=M(\xi,\eta,\alpha,m):$
\begin{eqnarray}
 M=\begin{pmatrix}
M_{11} & M_{12} & M_{13}  \\ 
M_{12} & M_{22}& M_{23} \\ 
M_{13}  & M_{23}& M_{33}
\end{pmatrix} m_{0},
\end{eqnarray}
Where,
\begin{eqnarray}\nonumber
M_{11}&=&\xi\eta^{2}+(1-\xi\eta^{2})\alpha^{2},\nonumber \\
M_{12}&=&-\sqrt{\frac{1-\alpha^{2}}{2}}(\xi\eta\sqrt{1-\eta^{2}})+\alpha(1-\xi\eta^{2})),\nonumber \\
M_{13}&=&-\sqrt{\frac{1-\alpha^{2}}{2}}(\xi\eta\sqrt{1-\eta^{2}})-\alpha(1-\xi\eta^{2})),\nonumber \\
M_{22}&=&\dfrac{1}{2}(1-\alpha^{2}+\xi(-\alpha\eta+\sqrt{1-\eta^{2}})^{2}),\nonumber \\
M_{23}&=&\dfrac{1}{2}(-1+\xi(1-\eta^{2})+\alpha^{2}(1-\xi\eta^{2}))\nonumber, \\
M_{33}&=&\dfrac{1}{2}(1-\alpha^{2}+\xi(\alpha\eta+\sqrt{1-\eta^{2}})^{2}).\nonumber \\
\end{eqnarray}
having eigenvalues\begin{equation}
0 ,\quad  m_{0}\xi\quad\text{and} \quad m_{0} 
\end{equation}
and solar mixing angle,
\begin{footnotesize}
\begin{eqnarray}
\tan2\theta_{12}=\dfrac{4\xi\eta\sqrt{1-\eta^{2}}(\sqrt{1-\alpha^{2}}\cos\theta_{13}+\alpha\sin\theta_{13})}{-1+2\xi-3\xi\eta^{2}+(-1+2\alpha^{2})(-1+\xi\eta^{2})\cos2\theta_{13}-2\alpha\sqrt{1-\alpha^{2}}(-1+\xi\eta^{2})\sin2\theta_{13}}.
\end{eqnarray}
\end{footnotesize}
In this case mass eigenvalues depend on the variables $(m_{0},\xi)$ whereas the solar mixing angle $\theta_{12}$ depends on the variables $(\xi,\eta,\alpha,\theta_{13})$ in a complicated way. 
\subsubsection{The unknown quantities : the absolute masses and the mixing angles}
For both cases disscussed above, the absolute masses are the three eigenvalues i.e.,
$m_{1}=0$, $m_{2}=m_{0}\xi$ and $m_{3}=m_{0}$. The corresponding eigenvectors  and the diagonalizing matrix are,
\begin{equation}\nonumber
U_{i}=\begin{pmatrix}
u_{i1} \\ 
u_{i2}\\ 
u_{i3}
\end{pmatrix}\hspace{1.3mm}
\text{and}\hspace{1.3mm}
U=\begin{pmatrix}
 u_{11}&u_{12}  &u_{13}  \\ 
  u_{21}& u_{22} &  u_{23}\\ 
u_{31} & u_{32} & u_{33}
\end{pmatrix} 
\end{equation}
respectively. Here, $U$ is identified as PMNS mixing matrix,
which leads to three mixing parameters, $
\tan^{2}\theta_{12}=(u_{12}/u_{11})^{2}$, $\tan^{2}\theta_{23}=(u_{23}/u_{33})^{2}$ and $\sin^{2}\theta_{13}=(u_{13})^{2}$.
\subsection{Numerical results and discussion}
On the basis of present available observational data, we can define certain interval for the parameters present in the matrices,
\begin{eqnarray}
m_{0}=[\hspace{1.3mm} 0.05004,\hspace{1.3mm} 0.05184\hspace{1.3mm}] eV \hspace{1.3mm}\text{and}\hspace{1.3mm}
\xi =[\hspace{1.3mm} 0.16625,\hspace{1.3mm} 0.17659\hspace{1.3mm}].
\end{eqnarray}
We see that within this range if we choose any value of $m_{0}$ and $\xi$ randomly, the values of $\Delta m^{2}_{21}$ and  $\Delta m^{2}_{32}$ are well fitted in the experimental $\pm1\sigma$ range. These are shown graphically in Figures 1- 2.
\begin{figure}
\begin{center}
\includegraphics[scale=1.4]{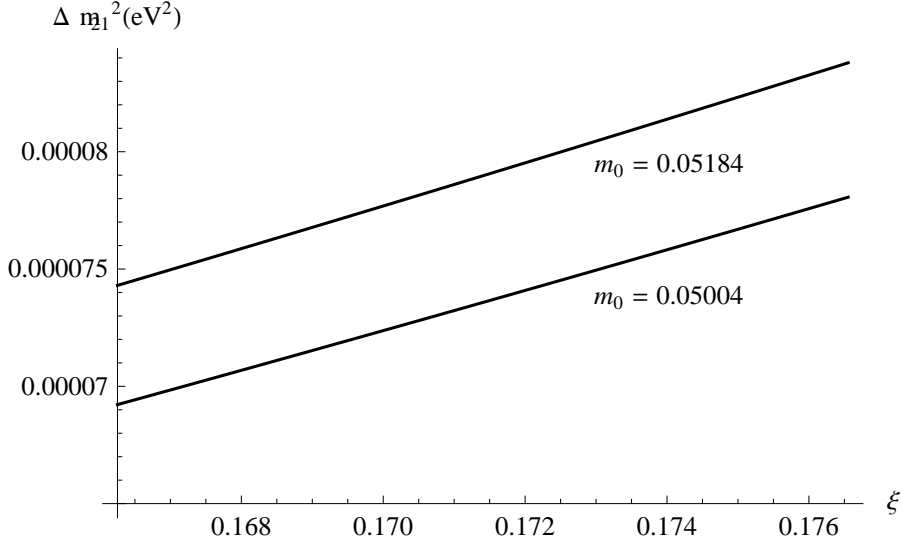} 
\caption{\footnotesize The variation of $\Delta m^{2}_{21}$ with  $\xi$ (N.H,for both $\theta_{13}=0$  and $\theta_{13}\neq0$) at $m_{0}= 0.05004$ $eV$  and $m_{0}= 0.05184$ $eV$. $\xi$ varies from 0.16625 to 0.17659.}
\end{center}
\end{figure}
\begin{figure}
\begin{center}
\includegraphics[scale=1.4]{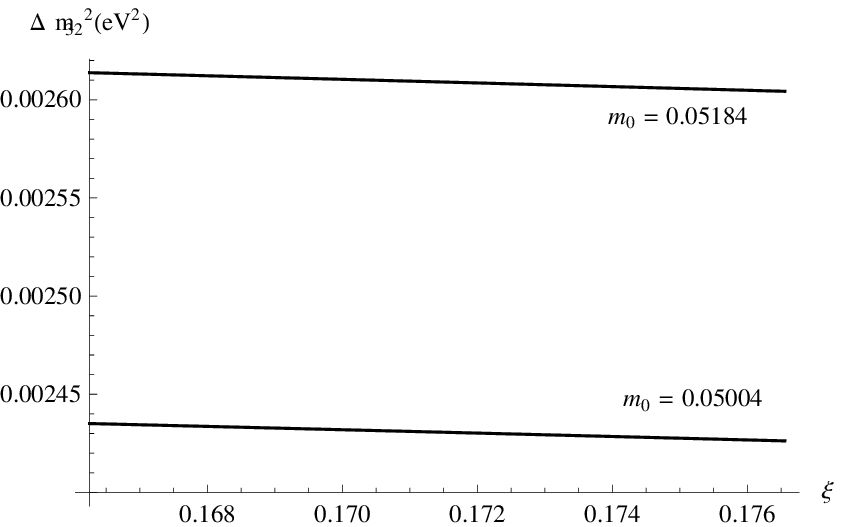} 
\caption{\footnotesize The variation of $\Delta m^{2}_{32}$ with  $\xi$ (N.H,for both $\theta_{13}=0$  and $\theta_{13}\neq0$) at $m_{0}= 0.05004$ $eV$  and $m_{0}= 0.05184$ $eV$. $\xi$ varies from 0.16625 to 0.17659.}
\end{center}
\end{figure}
\begin{figure}
\begin{center}
\includegraphics[scale=1.2]{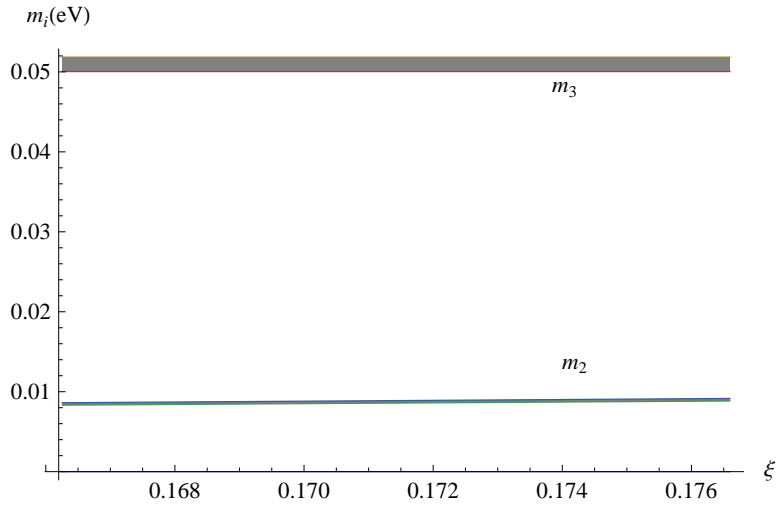} 
\caption{\footnotesize The variation of the three absolute masses w.r.t $\xi$ (N.H, for both $\theta_{13}=0$  and $\theta_{13}\neq0$) at $m_{0}=0.05004$ $eV$ and 0.05184 $eV$, and range of $\xi$ is between 0.16625 to 0.17659.}
\end{center}
\end{figure}
\begin{figure}
\begin{center}
\includegraphics[scale=1.2]{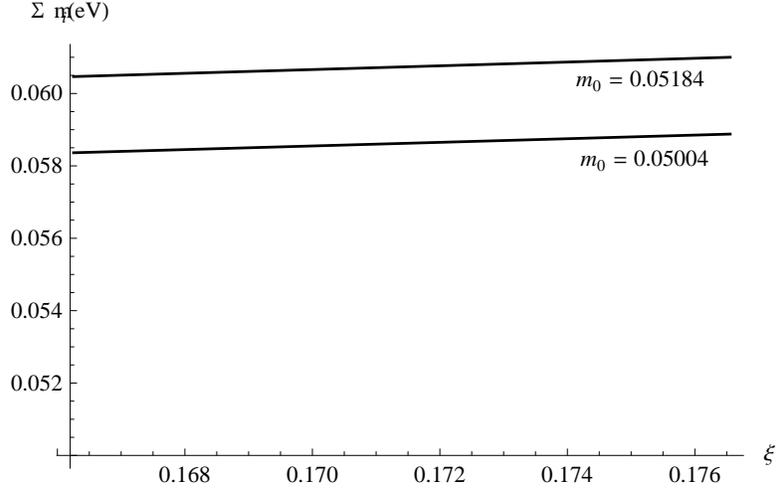} 
\caption{\footnotesize The variation of $\Sigma m_{i}$ w.r.t $\xi$ (N.H, for both $\theta_{13}=0$  and $\theta_{13}\neq0$)at $m_{0}=0.05004$ $eV$ and 0.05184 $eV$, and range of $\xi$ is between 0.16625 to 0.17659.}
\end{center}
\end{figure}

If we choose values for $m_{0}$ and $\xi$ as 0.05100 and 0.17335 respectively then we get,\begin{eqnarray}
m_{1} = 0,\quad m_{2}= 0.00884\hspace{1mm} eV \quad \text{and} \quad m_{3}=0.0510\hspace{1mm} eV.
\end{eqnarray}
leading to the observational parameters,
\begin{equation}
\Delta m^{2}_{21}=7.815\times10^{-5}\hspace{1mm}eV^{2},\hspace{1mm}
\Delta m^{2}_{32}=2.522\times10^{-3}\hspace{1mm}eV^{2}\hspace{1mm}\text{and}\hspace{1mm}
\Sigma m_{i}=0.0598\hspace{1mm}eV.
\end{equation}
The value of $\Sigma m_{i}$ is below the upper cosmologigal bound;$ < 0.28$ $eV$ [12].
Once we fix the values of $\xi$ and $m_{0}$, we can show TBM mixing and deviation by varying $\eta$ simply. Variation of $\eta$ does not affect the mass eigenvalues in both cases. Similarly, specific choices of values of $\xi$ and $m_{0}$ do not affect $\tan^{2}\theta_{12}$ in any way. We define the range of $\eta$, \begin{equation}
\eta = [\hspace{1mm}0.55045,\hspace{1mm}0.57879\hspace{1mm}].
\end{equation}
\begin{figure}
\includegraphics[scale=1.2]{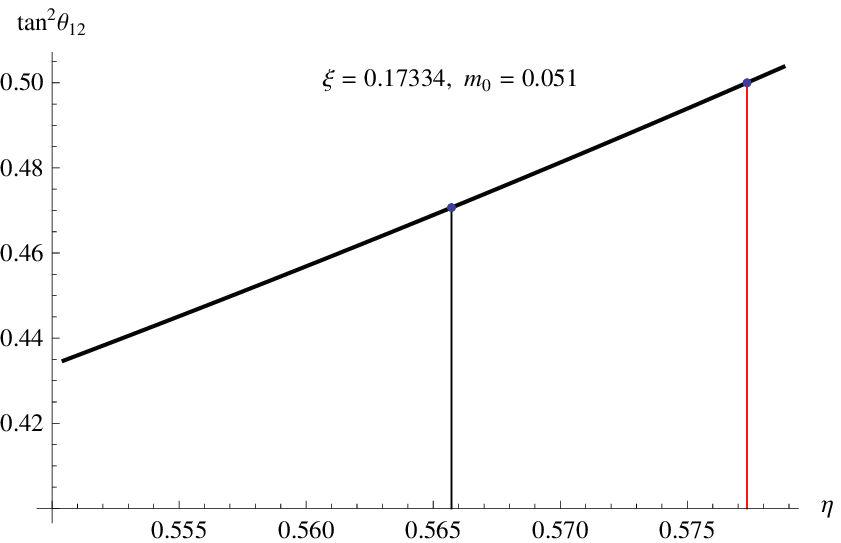}   
\caption{\footnotesize The variation of $\tan^{2}\theta_{12}$ against $\eta$ (N.H). The T.B.M is obtained at $\eta=0.57735$ (shown by red line) and the present experimental best fit value i.e $\tan^{2}\theta_{12}=0.47$ is obtained at about $\eta=0.5655$.}
\end{figure}
\begin{figure}
\includegraphics[scale=1.2]{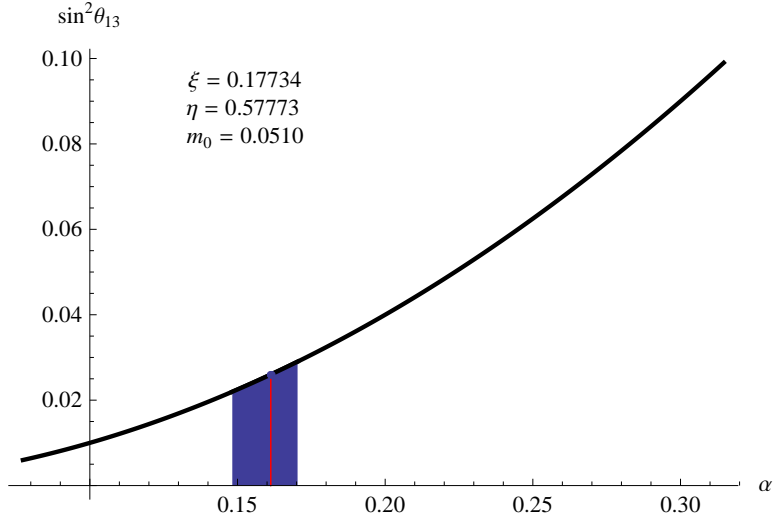}   
\caption{\footnotesize The variation of $\sin^{2}\theta_{13}$ against $\alpha$ (N.H). $\sin^{2}\theta_{13}=0.026$ is obtained at $\alpha=0.16124$ (red line).}
\end{figure}
\begin{figure}
\includegraphics[scale=1.2]{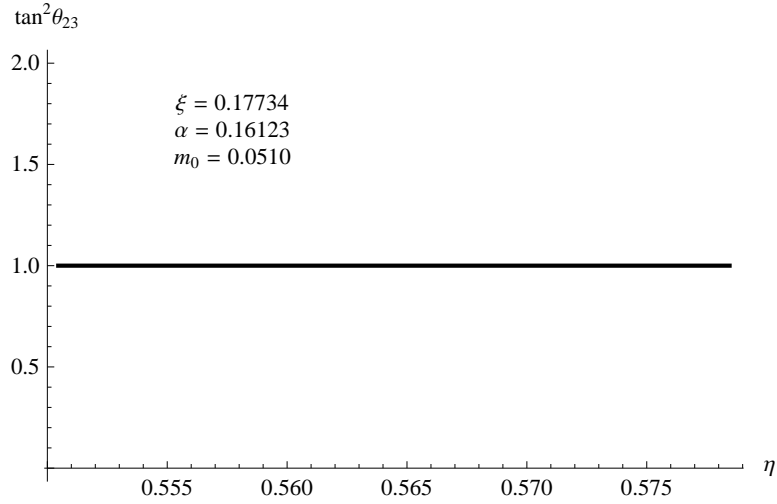}   
\caption{\footnotesize The variation of $\tan^{2}\theta_{23}$ against $\eta$ (N.H).}
\end{figure}
\begin{figure}
\includegraphics[scale=1.2]{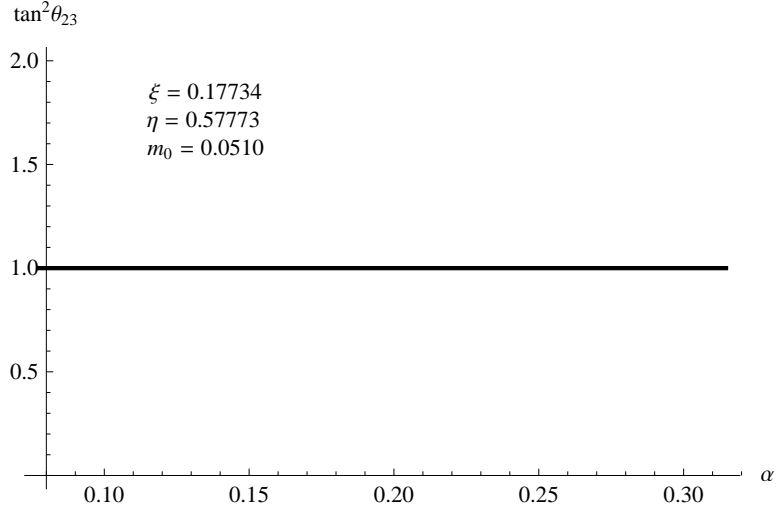}   
\caption{\footnotesize The variation of $\tan^{2}\theta_{23}$ against $\alpha$ (N.H).}
\end{figure}
Following the procedure described in Section 3.1.3, the variation of $\tan^{2}\theta_{12}$ w.r.t $\eta$ is shown graphically in Figure 5. Choosing $\eta = 0.57735$ leads to $\tan^{2}\theta_{12}=0.50$, whereas $\tan^{2}\theta_{12}=0.47$ (bestfit value) is obtained at $\eta=0.5655$.
Now if we consider the $\mu-\tau$ symmetric mass matrix of Eq.(20), we get $\sin^{2}\theta_{13} = 0$ all the way.
But along with the same ranges of $m_{0}$, $\xi$, $\eta$ and with the choosen values of $\xi$ and $m_{0}$, we can enter into the mass matrix $M$ of (30) with broken $\mu-\tau$ symmetry where the range of the new independent parametrer $\alpha$, that controls $\theta_{13}$, is defined under interval as,
\begin{equation}
\alpha = [\hspace{1mm}0.07733,\hspace{1mm}0.31463\hspace{1mm}].
\end{equation}
The variation of $\sin^{2}\theta_{13}$ is plotted against $\alpha$ in Figure 6 where we can further compress the range of $\alpha$ to $[\hspace{1mm}0.14823,\hspace{1mm}0.17033\hspace{1mm}]$ in order to comply with the experimental results within $\pm1\sigma$ range. The bestfit value of $\sin^{2}\theta_{13}=0.026$ is obtained at $\alpha=0.16123$.

We have shown graphically in Figures.7-8, how $\tan^{2}\theta_{23}$ maintains the maximality condition with change in $\alpha$ and $\eta$.
\section{Parameterisation of mass matrix for generating $\theta_{13}\neq0$, $\theta_{23}\neq \pi/4$ through a new perturbative matrix  $M_{s}$}
The mass matrix in Eq.(30) is capable of predicting the desired experimental results regarding $\tan^{2}\theta_{12}$ and $\sin^{2}\theta_{13}$, and at the same time satisfying the maximality condition of $\tan^{2}\theta_{23}$. Although the two guiding parameters $\alpha$ and $\eta$ are able to control independently the two mixing angles $\theta_{13}$ and $\theta_{12}$, yet this is contrary to some of the theoretical results [4, 6] that indicate dependency of $\tan^{2}\theta_{12}$ on $\sin^{2}\theta_{13}$ upto certain extent. So this is our honest intention to supress two of the parameters under a single one. We expect that this kind of supression of parameters will certainly affect the mass eigenvalues and the maximality condition of $\tan^{2}\theta_{23}$ upto certain extent. We now try to keep this effect as much small as posiible.
\subsection{The texture of perturbating mass matrix $M_{s}$}
Our procedure of parameterisation is as follows. We take the $\mu-\tau$ symmetric matrix that supports the TBM as the guiding matrix. Choice of TBM case as the reference one is supported by the fact that it can be related with certain symmetries. Our purpose is to derive out the matrix that can perturb this special case with minimum disturbance. The representation is something like,
\begin{equation}
M_{TBM} + M_{s}\longrightarrow M.
\end{equation}
Where $M_{s}$ differs from Eq.(25) and assumes the following symmetric structure [13,14],
\begin{eqnarray}
M_{s}=\begin{pmatrix}
r & y & -y \\ 
y & -x & z \\ 
-y & z & x'
\end{pmatrix}
\end{eqnarray}
Comparing the differences $M (\xi,m_{0},\eta,\alpha)-M_{\mu\tau}(\xi,m_{0},\eta_{0})$, with $\eta_{0}$ being fixed at TBM mixing, we get an idea about the range of numerical values that $r$, $x$, $x'$, $z$ and $y$ can take, by applying the intervals of $\eta$ and $\alpha$ specified earlier in Eqs.(37) and (38). On the basis of that  we choose certain numerical values for all the members present in $M_{s}$, except for $y$  which we generalize to $\omega$, as one parameter.
From our study, $M_{s}$ assumes the following emperical structure.
\begin{equation}
M_{s}=\begin{pmatrix}
 0.2& 1  & -1  \\ 
1 & -0.1 & 0.05 \\ 
-1 & 0.05 & 0.009
\end{pmatrix}\omega ,
\end{equation}
The parameter $\omega$ not only dictates the deviation from TBM and $\theta_{13}$ but also devaiates $\tan^{2}\theta_{23}$ from unity as well. Choosing a wide range of $\omega$, we can produce a series of  of mass matrices with broken $\mu-\tau$ symmetry, and hence can generate a series of values regarding $\tan^{2}\theta_{12}$, $\sin^{2}\theta_{13}$, $\tan^{2}\theta_{23}$, $\Delta m^{2}_{21}$, $\Delta m^{2}_{32}$ and $\Sigma m_{i}$ against $\omega$. We define a bound on the the parameter $\omega$ as $[\omega_{1},\omega_{2}]$ based on the experimental $\pm1\sigma$ range of $\tan^{2}\theta_{12}$. We define $\omega=\omega_{0}$ corresponding to the bestfit value of $\tan^{2}\theta_{12}$. Our purpose is to check the values of all the quantities stated above within this range, to point out the values of the specific quantities at $\omega=\omega_{0}$ and to compare with present experimental results. From Eqs.(20) and (41), $M_{NH}$ in Eq.(39) becomes,

\begin{scriptsize}
\begin{eqnarray}
M_{NH} = \begin{pmatrix}
 \xi \eta^{2}& - \xi \eta \sqrt{\frac{1-\eta^{2}}{2}}  & - \xi \eta \sqrt{\frac{1-\eta^{2}}{2}} \\ 
- \xi \eta \sqrt{\frac{1-\eta^{2}}{2}} & \frac{1}{2}(1 + \xi (1- \eta^{2})) & \frac{1}{2}(-1 + \xi (1- \eta^{2})) \\ 
- \xi \eta \sqrt{\frac{1-\eta^{2}}{2}} &\frac{1}{2}(-1 + \xi (1- \eta^{2}))  & \frac{1}{2}(1 + \xi (1- \eta^{2}))
\end{pmatrix} m_{0} +\begin{pmatrix}
 0.2& 1  & -1  \\ 
1 & -0.1 & 0.05 \\ 
-1 & 0.05 & 0.009
\end{pmatrix}\omega .
\end{eqnarray}
\end{scriptsize}
We fix the values of the parameters except $\omega$ as follows.
For N.H,
\begin{eqnarray}\nonumber
\xi =0.173346,\hspace{1.2mm}
\eta= 0.57735,\hspace{1.2mm}\text{and}\hspace{1.2mm}
m_{0} = 0.0510 eV.
\end{eqnarray}
When $\omega=0$, with values of the parameters prescribed above, T.B.M can be generated.  
\subsection{Numerical results for Normal Hierarchy (N.H) using perturbative matrix $M_{s}$}
We consider the following range of input parameters for numerical analysis.
 For N.H,
\begin{eqnarray}
[\hspace{1mm}\omega_{1},\hspace{1mm}\omega_{2} \hspace{1mm}]=[\hspace{1mm}0.00408,  \hspace{1mm}0.00637\hspace{1mm}]\hspace{1.4mm}\text{and}\hspace{1.4mm}
\omega_{0}=0.00534.
\end{eqnarray}
The detailed results are shown graphically and the summary is presented in the form of Table 1 for Normal Hierarchy case. We have compared the results against $\omega_{0}$ and compared them against the bestfit results from  recent experimental data. In order to include the precision of the results we define a range $[\omega_{1},\omega_{2}]$, and against that boundary the parameters like $\tan^{2}\theta_{12}$, $\sin^{2}\theta_{13}$, $\Delta m^{2}_{21}$ etc. assume certain ranges. We compare that ranges with the experimental $1\sigma$, $2\sigma$ and $3\sigma$ boundaries [3].

From graphs in the Figures $9-10$, the variation of $\tan^{2}\theta_{12}$ against $\sin^{2}\theta_{13}$ can be studied. It can be seen that within the interval [$\omega_{1}$, $\omega_{2}$] the increase in $\sin^{2}\theta_{13}$ is associated with decrease in $\tan^{2}\theta_{12}$ and slight deviation of $\tan^{2}\theta_{23}$ from unity (Figure 11). The variation of $\Delta m_{21}^{2}$ and $\Delta m_{32}^{2}$ with $\omega$ are plotted in Figures $12-13$ respectively. The variation of $\Sigma m_{i}$ w.r.t $\omega$ has also been plotted in Figure 14, which is also found to be within the upper cosmological bound, i.e., $\Sigma m_{i}\leqslant0.28$ $eV$.
\begin{table}
\begin{center}
\begin{tabular}{|c|c|c||c|c|c|}
\hline parameter & $\omega_{0}$ &[$\omega_{1},\omega_{2}$]  & bestfit$\pm $ $1\sigma$ & $2\sigma$ & $3\sigma$  \\ 
\hline \hline $\tan^{2}\theta_{12}$ & $0.47 $& $[0.435, 0.503]$ & $0.47^{+0.033}_{-0.036}$ &$0.408 -0.538$  & $0.370-0.587$ \\ 
 $\sin^{2}\theta_{13}$ &$0.025$  & $[0.015, 0.035]$ & $0.026^{+0.003}_{-0.004}$ & $0.019-0.033$ & $0.015-0.036$ \\ 
 $\tan^{2}\theta_{23} $& $0.919 $& $[0.937, 0.906] $ & $ 0.96^{.364}_{-.175}$ & $0.695-1.631$ & $0.639-1.778$ \\ 
 $\Delta m^{2}_{21} (10^{-5} eV^{2})$ &$7.587 $ & $[7.357, 7.794 ]$ &$7.62 \pm 0.19$  & $7.27-8.01$ & $7.12-8.20 $ \\ 
 $\Delta m^{2}_{32} (10^{-3} eV^{2})$ &$2.598 $ & $[2.556, 2.642]$& $2.53^{+0.08}_{-0.10}$ & $2.34-2.69$ & $2.26-2.77$ \\ 
\hline 
\end{tabular} 
\caption{\footnotesize The Neutrino oscillation parameters (N.H) for $\omega_{0}$ and [$\omega_{1}$,$\omega_{2}$], and comparing them with recent experimental results[3]. For N.H, $\omega_{0}=0.00534$, $\omega_{1}=0.00408$ and $\omega_{2}=0.00637$.}
\end{center}
\end{table}
 \begin{figure}
 \includegraphics[scale=1.4]{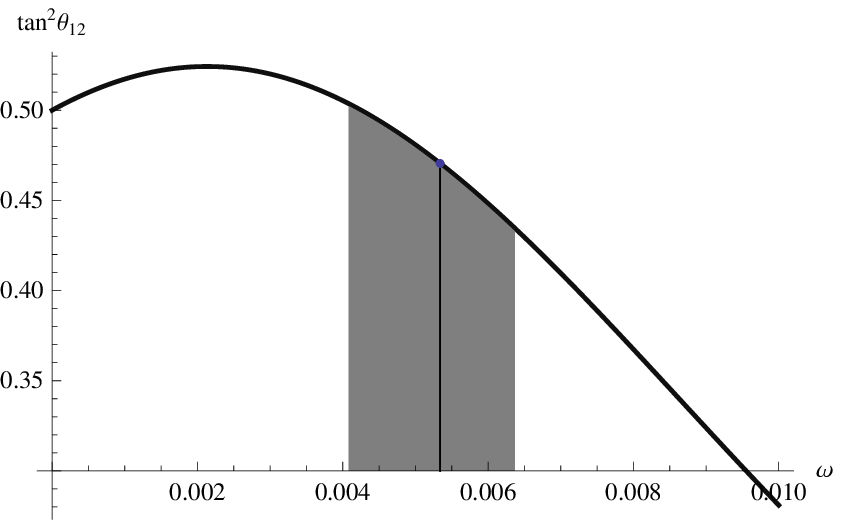}
\caption{\footnotesize The variation of $\tan^{2}\theta_{12}$ w.r.t $\omega$ (N.H).}
\end{figure}
\begin{figure}
 \includegraphics[scale=1.4]{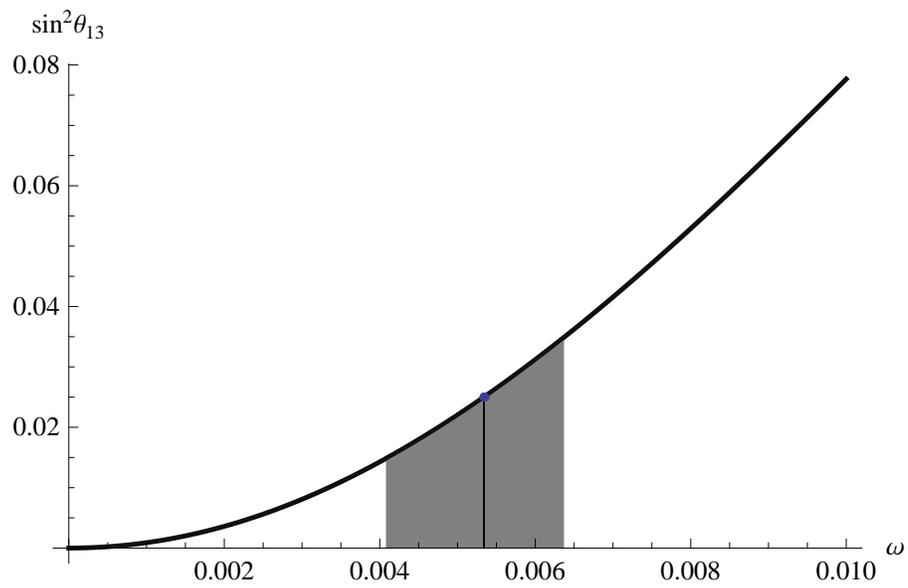}
\caption{\footnotesize The variation of $\sin^{2}\theta_{13}$ w.r.t $\omega$ (N.H).}
\end{figure}
\begin{figure}
\includegraphics[scale=1.4]{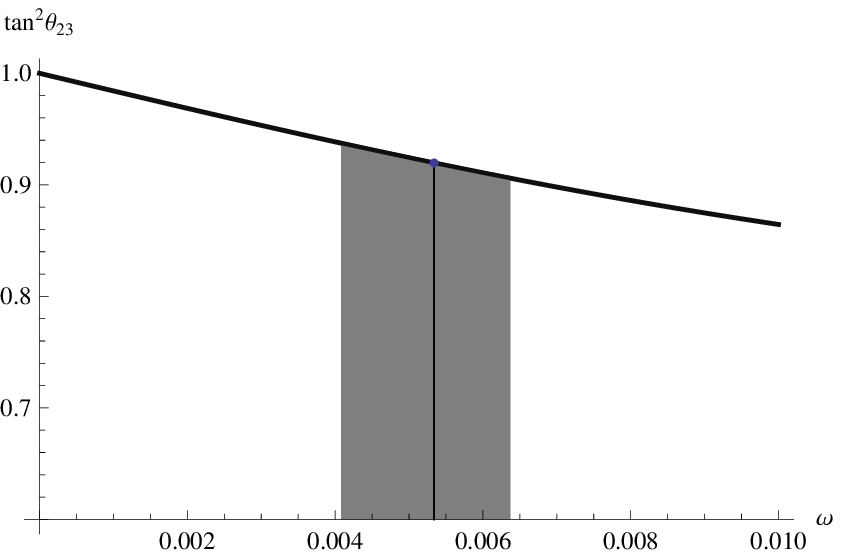} 
\caption{\footnotesize The variation of $\tan^2\theta_{23}$ w.r.t $\omega$ (N.H).}
\end{figure}
\begin{figure}
\includegraphics[scale=1.4]{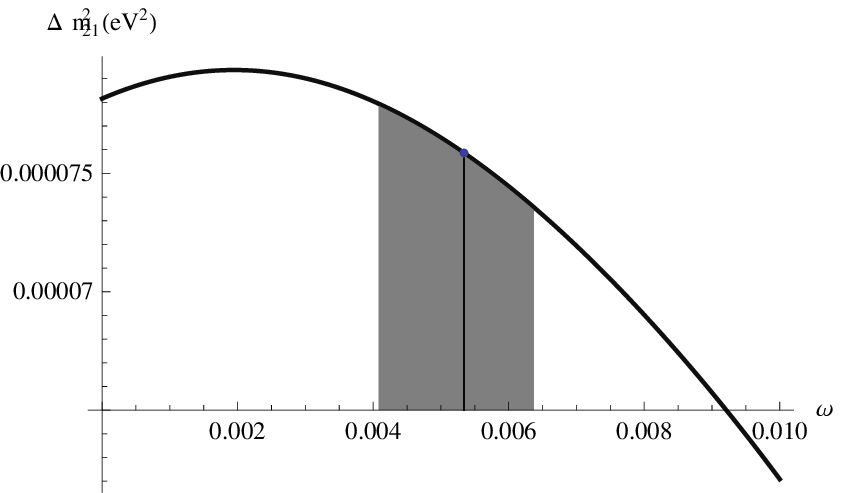} 
\caption{\footnotesize The variation of $\Delta m^{2}_{21}$ w.r.t $\omega$ (N.H).}
\end{figure}
\begin{figure}
\includegraphics[scale=1.4]{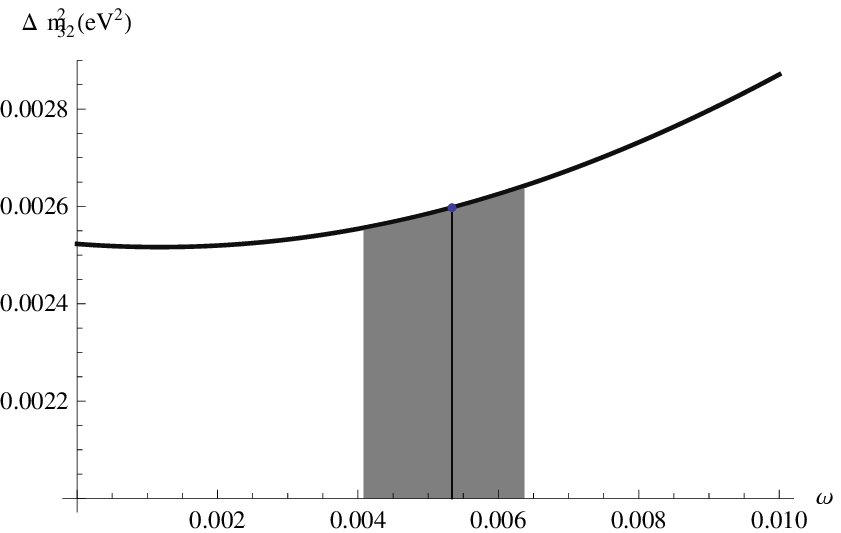} 
\caption{\footnotesize The variation of $\Delta m^{2}_{32}$ w.r.t $\omega$ (N.H).}
\end{figure}
\begin{figure}
\includegraphics[scale=1.2]{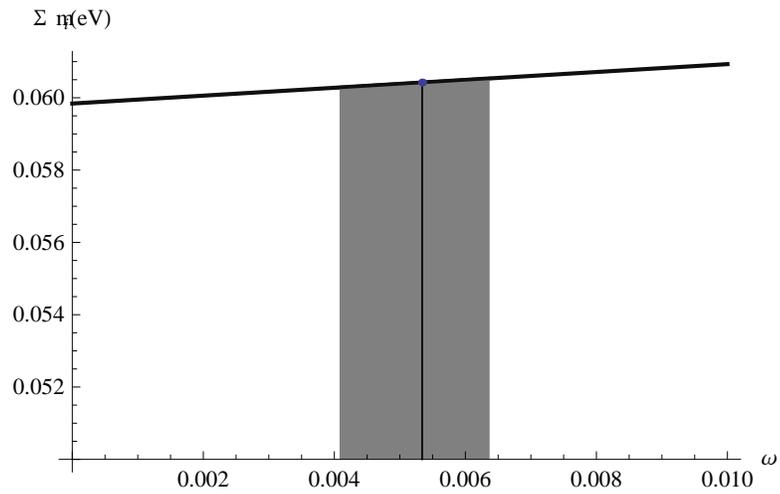} 
\caption{\footnotesize The variation of $\Sigma m_{i}$ w.r.t $\omega$ (N.H).}
\end{figure}
\section{Summary and discussions}
We stared with the objective that the mass matrix must involve eigenvalues having simple structure, and should not involve the controlling parameters of $\theta_{12}$ and $\theta_{13}$. We have choosen three  parameters ($\xi$, $\eta$, $m_{0}$) if the mass matrix is $\mu-\tau$ symmetric, and another additional parameter $\alpha$ when the $\mu-\tau$ symmetry is broken, for the parameterisation of the mass matrix, with an assumption $\delta_{cp}=0$. The stability of the mass eigenvalues and the maximality condition of $\tan^{2}\theta_{23}$ with change in $\theta_{12}$ and $\theta_{13}$, are the special features of this method. The parameterisation is also characterized by freedom of choosing the numerical values of the independent parameters and to adjust itself with any change. In the next part we have tried to visualize the perturbation with respect the $\mu-\tau$ symmetric mass matrix satisfying TBM mixing. We have tried to restrict the degrees of freedom of the mass matrix proposed earlier, by introducing a new  emperical perturbing symmetric mass matrix $M_{s}$ with one single parameter $\omega$. The texture of $M_{s}$ is constructed on the basis of the  differences of $M$ and $M_{\mu\tau}$. The parameter $\omega$ helps to study the variation of $\tan^{2}\theta_{12}$ w.r.t change in $\sin^{2}\theta_{13}$ and  at the same time deviation of $\tan^{2}\theta_{23}$ from unity as well. We investigated results of all the oscillation parameters at certain preferred value $\omega_{0}$ and are found very close to the experimental bestfit or within the experimental $\pm 1\sigma$ range. In addition, we have also defined a range $[\omega_{1},\omega_{2}]$ and studied the oscillation parameters within that interval and compared with the $2\sigma$ and $3\sigma$ range.
The generalization of the present method of parameterisation to quasidegenerate case [15] with normal hierarchy (QD-NH) is in progress with great success. The same can be extended to inverted case [9] as well.  
\section*{Acknowledgement}
One of the author, S Roy conveys his heartfull gratitude to Rituparna Chutia, from Department of Staticstics, Gauhati University for his kind assistance regarding interval analysis.  
\bibliographystyle{plain}

\end{document}